\documentclass[review]{elsarticle}

\usepackage{hyperref}
\usepackage{url}
\usepackage{booktabs}
\usepackage{subfig}
\usepackage{graphicx}
\usepackage{float}
\usepackage[cmex10]{amsmath}
\usepackage{multirow}

\begin{document}

\begin{frontmatter}

\title{Extroverts Tweet Differently from Introverts in Weibo}
% \tnotetext[mytitlenote]{Fully documented templates are available in the elsarticle package on \href{http://www.ctan.org/tex-archive/macros/latex/contrib/elsarticle}{CTAN}.}n

% Group authors per affiliation:
\author{Zhenkun Zhou, Ke Xu}
\address{State Key Lab of Software Development Environment, Beihang University, Beijing, China}
\author{Jichang Zhao*}
\address{School of Economics and Management, Beihang University, Beijing, China\\
$^*$Corresponding author: jichang@buaa.edu.cn
}

\begin{abstract}

Being dominant factors driving the human actions, personalities can be excellent indicators in predicting the offline and online behavior of different individuals. However, because of the great expense and inevitable subjectivity in questionnaires and surveys, it is challenging for conventional studies to explore the connection between personality and behavior and gain insights in the context of large amount individuals. Considering the more and more important role of the online social media in daily communications, we argue that the footprint of massive individuals, like tweets in Weibo, can be the inspiring proxy to infer the personality and further understand its functions in shaping the online human behavior. In this study, a map from self-reports of personalities to online profiles of 293 active users in Weibo is established to train a competent machine learning model, which then successfully identifies over 7,000 users as extroverts or introverts. Systematical comparisons from perspectives of tempo-spatial patterns, online activities, emotion expressions and attitudes to virtual honor surprisingly disclose that the extrovert indeed behaves differently from the introvert in Weibo. Our findings provide solid evidence to justify the methodology of employing machine learning to objectively study personalities of massive individuals and shed lights on applications of probing personalities and corresponding behaviors solely through online profiles. 

\end{abstract}


\end{frontmatter}

%\linenumbers

\section{Introduction}
\label{introduction}

The online social media has being becoming an essential component of everyday life, which even reflects all aspects of human behavior. Millions of users have digitalized and virtualized themselves in popular platforms like Twitter and Weibo, including basic demographics, plenty of statuses, abundant emotions and diverse activities. These online profiles can be natural, detailed, long-term and objective footprints of massive individuals and thus they could be promising proxies in understanding human personalities~\cite{Boyd:2007eu,Barash:2010}. Since its beginning being a sub-discipline of psychology, the study of human personalities has aimed at one general goal, which is to describe and explain the significant psychological differences between individuals. Revealing the connection between different personalities and corresponding behavioral patterns, especially in the circumstance of online social media, is one of the most exciting issues~\cite{Maddi:1980,Mcadams:1990,Larsen:2008} in recent decades. And a growing body of evidence implying individual personality discrepancy in online social media further makes it imperative in probing online human behavior from views of personalities~\cite{Ross:2009gp,Ryan:2011cq,Simoncic:2014}.

% Fault of method
Personality is a stable set of characteristics and tendencies which specify similarities and differences in individuals' psychological behavior and it is also a dominant factor in shaping human thoughts, feelings and actions. However, personality traits, like many other psychological dimensions, are latent and hard to be measured directly. Self-report of asking subjects to fill survey questionnaires referring to personalities is a classical way to assess respondents in the conventional studies~\cite{Costa:1992,ROBERTS:2005in,Fast:2008}, while its limitations are inevitable and can be summarized as: 
\begin{itemize}
\item Expensiveness. Questionnaires in self-reports can be much time-consuming and costly and even worse, the response rate might be unexpectedly low~\cite{hoonakker2009questionnaire} and all these concerns will badly reduce the valid number of participants, which is generally below 1,000~\cite{Watt2004Internet}. And it is challenging to come to persuasive and universal conclusions based on such a small number of samples. 

\item Subjectivity. Respondents fill in the questionnaires mainly based on their cognition, memory or feelings, and they could hide the true responses or thoughts consciously or unconsciously while facing the questions. Particularly for self-reports referring personalities, they might even not recollect the circumstance exactly in the controlled lab environments. 

\item Low flexibility. Questionnaires are generally designed according to the study assumptions before conducting the experiments and it is hard to obtain insights that out of the scope of the previously established goals, i.e., existing self-reports might be much less inspiring because of lacking extension.
\end{itemize}

To some extent, the above limitations can be overcome because of the emergence of crowdsourcing marketplaces like Amazon Mechanical Turk (MTurk), which offer many practical advantages that reduce costs and make massive recruitments feasible~\cite{Paolacci:2010running} and become dominant sources of experimental data for social scientists. While in the meantime, new concerns are brought in~\cite{wright2005researching,Bohannon:2016}. For example, researchers concern that the volunteers are less numerous and diverse than their hope, while Turkers complain that the reward is too low. In addition, MTurk has suffered from the growing participant non-naivety~\cite{peer2017beyond}. Accounting for these shortages, the recent progress in machine learning, especially the idea of computation driven solutions in social sciences~\cite{Lazer721}, shows an increasing interest in modeling and understanding of human behavior such as personalities.

% New opportunities
Indeed, the popularity of online social media provides a great opportunity to examine personality inference using significant amounts of data. Taking Weibo as an example, about 100 million Chinese tweets are posted everyday and from which we can sense the online behavior of 500 million users of tremendously diverse backgrounds. 
The development report from Weibo in 2015 officially shows the number of monthly active users is around 222 million. 
These numbers imply further that the availability of vast and rich datasets of active individuals' digital fingerprints from online social media will unprecedentedly increase the scale and granularity in measuring and understanding human behavior, especially for personalities, because the cost of the experiment will be essential reduced, the objectivity of the samples will be convincingly guaranteed and the flexibility of the data will be adequately amplified.
At the same time, there are new opportunities to combine social media with traditional surveys in personality psychology. Kosinski et al. demonstrate that available digital records in Facebook can be used to automatically and accurately predict personalities~\cite{Kosinski:2013gi}. With the help of developments in machine learning, computer models can make valid personality prediction, even outperform the self-reported personalty scores~\cite{Youyou:2015iu}. In this study, we argue that from the perspective of computational social science, profiles of active users in Weibo can be excellent proxies in probing the interplay between personalities and online behavior.

% My work
An online page with a 60-items version of the Big-Five Personality Inventory is established first in our study to collect scores on personality traits~\cite{McCrae:2004ie} and a total of 293 valid users in Weibo are asked to finish the self-report on this page, which provides a baseline for the following study. Focusing on extraversion, the scores mainly follow Gaussian distribution and the subjects are accordingly divided into three groups of high, neutral and low scores on extraversion. Then by collecting online profiles of those self-reporters from Weibo, a map between the self-reports of extraversion and the online profiles is built to train machine learning models that can automatically evaluate the extraversion of much more individuals without the help of self-reports. Three kinds of features, including 13 basic ones, 33 behavioral ones and 84 linguistic ones are comprehensively considered in the SVM model and its performance is also convincingly justified by cross-validations. With over 7,000 users being labeled as extroverts or introverts by the model, we attempt to systematically study the difference of online behavior caused by extraversion through investigating into the following seven research questions:

\textbf{RQ1}. Do extroverts and introverts tweet temporally differently in Weibo? 

\textbf{RQ2}. Do extroverts and introverts tweet spatially differently in Weibo?

\textbf{RQ3}. What types of information do extroverts and introverts prefer to share?

\textbf{RQ4}. Who is more socially active in the online circumstance, extroverts or introverts?

\textbf{RQ5}. Who pay more attention to online purchasing and shopping, extroverts or introverts?

\textbf{RQ6}. Do extroverts and introverts express emotions differently in Weibo?

\textbf{RQ7}. Who care more about the online virtual honor, extroverts or introverts?

According to these questions, unexpected differences in online behavior of extroverts and introverts are disclosed. Introverts post more frequently than extroverts, especially at the daytime. However, extroverts visit different cities instead of staying at just one familiar city as the introvert does. The spatial discrepancy can be more unintuitive as we zoom in to better resolutions, for example, introverts tend to check in themselves while shopping, however, extroverts enjoy posting at working places. In addition, a tiny fraction of introverts might attempt to camouflage their own loneliness to others by tweeting with a large number of different areas ($>20$). Extroverts enjoy sharing music and selfies while introverts prefer retweeting news. As to online interactions, extroverts mention friends more than introverts, implying higher social vibrancy. By presenting a purchasing index to depict the online buying intention, we find that as compared to the extrovert, introverts devote more efforts in posting shopping tweets to relieve the loneliness due to a lack of social interaction with others. We also categorize the emotion delivered in tweets into anger, disgust, happiness, sadness, and fear~\cite{Zhao:2012jc} and find that introverts post more angry and fearful (high arousal) tweets and extroverts post more sad (low arousal) ones. Finally, extroverts attach more meanings to the online virtual honor than introverts do, implying that they might be ideal candidates for online promoting campaigns with virtual honor. To our best knowledge, this is the first study to completely compare the online behavior of extroverts and introverts over large-scale samples and our findings will be helpful in understanding the role of personalities in shaping human behavior.

\section{Literature Review and Theoretical Background}

Several well studied models have been established for personality traits and in which Big-Five model is the most popular one~\cite{Goldberg:1992ce,Gosling:2003eh}. In this model, human personality can be depicted from five dimensions, including openness, neuroticism, extraversion, agreeableness and conscientiousness and the personality type could be identified through individual's behavior over the time and circumstances. The Internet, one of the most pervasive circumstance today, has in fact profoundly changed the human behavior and experience. With its explosive development, lots of research efforts have been devoted in investigating the relation between personality and Internet usage. For example, the findings of Amiel et al. demonstrate that distinctive patterns of Internet use and usage motives for those of different personality types and extroverts made more goal-oriented use of Internet services~\cite{Amiel:2004fz}. Focusing on online social media, as the vital component of the Internet, extraversion and openness to experiences are found to be positively related with social media adoptions~\cite{Correa:2010ea}.

In the meantime, it was also pointed out that users' psychological traits could be inferred through their digital fingerprints in online social media~\cite{Ellison:2007di,Ong:2011hx}. Golbeck et al. proposed to bridge the gap between personality study and social media and demonstrated that social media (Facebook and Twitter) profiles can reflect personality traits~\cite{Golbeck:2011kr,Golbeck:2011be}. They suggested that the number of parentheses used is negatively correlated with extraversion, however, explanations beyond the correlation is not provided and probing the correlations over a larger data set still remains necessary. Quercia et al. employed numbers of followees, followers and tweets to learn the personality and suggested that both popular users and influentials are extroverts with stable emotions~\cite{Quercia:2011iu}. Besides, patterns in language use of online social media, like words, phrases and topics also offer a way to reveal personalities~\cite{Schwartz:2013bt}. For example, using dimensionality reduction for the Facebook Likes of participants, Kosinski et al. proposed a model to predict individual psycho-demographic profiles~\cite{Kosinski:2013gi}. As for social media in China, Weibo and RenRen become the ideal platforms for conducting personality research~\cite{Li:2014cs,Bai:2012dq}. Considering the recent progress that computer algorithms outperform humans in personality judgment~\cite{Youyou:2015iu}, online social media indeed offer unprecedented opportunities for personality inferring and human behavior understanding. 

Each bipolar dimension (like extraversion) in Big-Five model summarizes several facets, which subsumes lots of more specific traits (extraversion vs. introversion). In this paper, we focus on the extraversion which is an indispensable dimension of personality traits. Many efforts from previous studies have been delivered to reveal the connection between extraversion and online behaviors and can be roughly reviewed from the following perspectives. 

\textbf{Social interactions} Highly extroverted individuals tend to have broad social communications with others~\cite{Li:2014cs}. For instance, extraversion generally positively related to the number of Facebook friends~\cite{Bachrach:2012,Amichai:2010}. Gosling et al. also found particularly strong consensus about Facebook profile-based personality assessment for extroverts~\cite{Gosling:2007}. However, Ross et al.~\cite{Ross:2009gp} showed that extroverts are not necessarily associated with more Facebook friends, which are contrary to later results of Bachrach et al.~\cite{Bachrach:2012} and Hamburger et al.~\cite{Amichai:2010}. Through posting tweets, extroverts are more actively sharing their lives and feelings with other people and the personality traits might shape the language styles in social media. In English, extroverts are more likely to mention social words such as `party' and `love you', whereas introverts are more likely to mention words related to solitary activities such as `computer' and `Internet'~\cite{Schwartz:2013bt}. Referring to Chinese, extraversion is positively correlated with personal pronouns, indicating that extroverts tend to be more concerned about others~\cite{Qiu:2016bx}.

\textbf{Buying intention} Extraversion, as one personality trait, is one of main factor in in driving online behaviors including buying, and hence exploring the relationship between extraversion and shopping is a valuable topic. DeSarbo and Edwards found that individuals of social isolation tend to perform compulsive buyings in efforts to relieve the feelings of loneliness due to a lack of interaction with others~\cite{DESARBO:1996ck}. However, the results of subsequent studies about the relationship between compulsive buying and extraversion are inconsistent~\cite{mowen1999understanding,Gohary:2014ba}.

\textbf{Emotion expression} In psychology, it is widely believed that extraversion is associated with higher positive affect, namely extroverts experience increased positive emotions~\cite{mccrae2003personality,Smillie:2015fl}. Extroverts are also more likely to utilize the supplementary entertainment services provided by social media, which bring them more happiness~\cite{Deng:2013if}. While, Qiu et al. suggested that highly extroverted participants do use it to relieve their existential anxiety in social media~\cite{Qiu:2010vw}. Thus, it is necessary to investigate the relation between various emotions and extraversion rather than only the positive affect.

However, most existing studies built their conclusions on self-reports from very small samples and the lacking of data or objectivity leads to inconsistent or even conflicting results. Moreover, a comprehensive understanding of how extroverts and introverts behave differently in the circumstance of online social media still remains unclear. Hence in this study, we try to employ machine learning models to identify and establish a large group of samples and then investigate the behavioral difference from diverse aspects, aiming at offering solid evidence and comprehensive views. 

\section{Identification of the extraversion}

\subsection{Dataset and participant population}

The Big-Five model is the most accepted and commonly used model in depicting human personalities~\cite{Goldberg:1992ce,Gosling:2003eh} and quite a few measuring instruments have also been developed to assess the Big-Five personality traits. In this study, a web page with a 60-question version of the Big-Five Personality Inventory~\cite{McCrae:2004ie} is built to collect self-reported scores on different personality traits. We target on Weibo users for voluntary participants recruitment and invitations were sent via both online and offline manners ranging from December 1, 2014 to March 31, 2015. All the participants are manually checked and only valid ones in Weibo (can be identified by the Weibo ID, a unique identification for each user) are considered. Finally a total of 293 valid participants are selected in the following study (144 men and 149 women) and the age of all participants ranges from 19 to 25. It is worth noting that according to the official report of Weibo in 2015, users with age between 17 to 33 occupy around 80\% of its population, indicating that our refined samples of self-reports can sufficiently represent the most users in Weibo.

\begin{figure}
\centering
\includegraphics[width=0.8\textwidth]{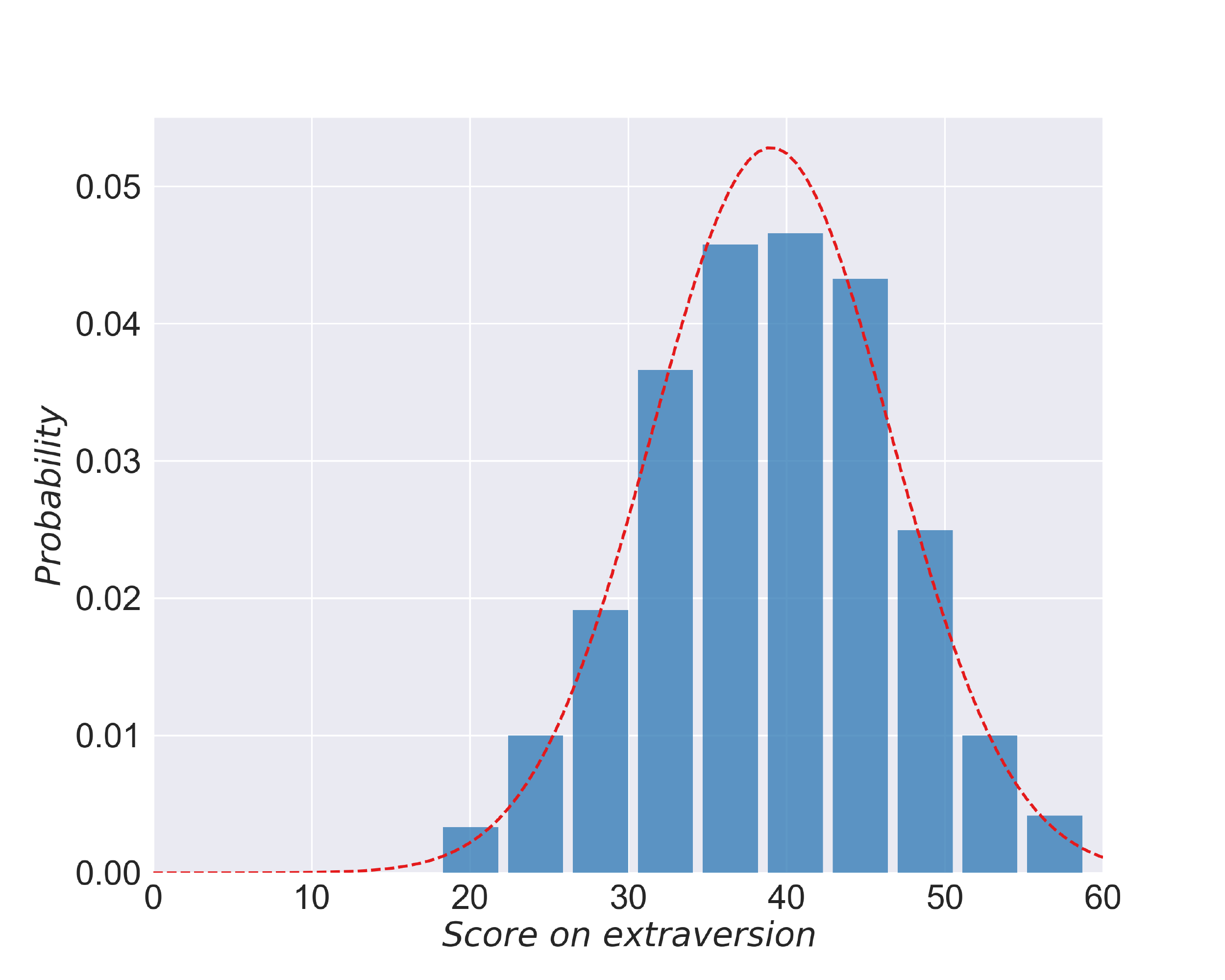}
\caption{The distribution of scores on extraversion from self-reports of 293 valid samples.}
\label{fig:ext_dist}
\end{figure}

We focus on the extraversion of Big-Five personality traits in this study, which measures a personal tendency to seek stimulation in the external world, company of others, and express positive emotions~\cite{Goldberg:1992ce}. People who score high in extraversion (called extroverts) are generally outgoing, energetic and friendly. On the contrary, introverts are more likely to be solitary and seek environments characterized by lower levels of external simulation. The distribution of scores from 293 valid samples (Weibo users) on extraversion is shown in Fig.~\ref{fig:ext_dist}. The scores follow a typical Gaussian distribution with $\mu$ (mean value) being 39.03 and $\sigma$ (standard deviation) being 7.55. It can be seen in Fig.~\ref{fig:ext_dist} that the probability of scores near the mean value is relatively higher than the occurrence of both high scores and low scores, implying that a significant fraction of samples report the neutral scores on extraversion and they can be intuitively categorized to the type of without much significantly distinct personality, i.e., neither extroverts nor introverts. Because of this, it is reasonable to divide samples into three groups including extroverts (with high scores and labeled as 1), neutrals (with scores around the mean and labeled as 0) and introverts (with low scores and labeled as -1). Specifically, extroverts are samples with scores more than 42.81 ($\mu+\sigma/2$), introverts are users with scores less than 35.25 ($\mu-\sigma/2$) and neutrals represent users whose scores ranging from 35.25 to 42.81. The thresholds ($\mu\pm\sigma/2$) are set to balance the size of three categories, aiming at avoid the bias in machine learning models. By labeling 293 valid samples into three categories, we can obtain a training set for establishing and evaluating machine learning models that do not need the help of self-reports.

With the permission granted by valid samples in the self-reports, we continuously collect their online profiles until March 1, 2016, including demographics and posted tweets through Weibo's open APIs. In order to guarantee the quality of the data, only users with more than 100 tweets are remained to build the training set, including 45 extroverts (1), 44 introverts (-1) and 56 neutrals (0). The training data is generally balanced on the three classification labels, especially for extroverts and introverts, which is helpful to avoid the bias of the machine learning model.

\subsection{Extraversion classifier}
\label{sec:clas}

As reviewed in the former section that many aspects of online profiles have been previously found to be connected with users' personalities, hence for the purpose of establishing a competent classifier to convincingly identify the three categories of extraversion without the help of self-reports, we try to extract as many features as we can from the digital and textual records and these features are roughly grouped into basic ones, interactive ones and linguistic ones. 
Details of different kinds of features are introduced as follows, respectively.

\textbf{Basic features} Basic features are selected to reflect the user's demographics, preliminary statuses and elementary interactions in the social media, including gender, tweeting patterns and privacy settings. 
Specifically, tweeting patterns contain $log(ARS+1)$ (where $ARS$ is the age of a user in Weibo since its register with unit of day), $log(NT+1)$ (where $NT$ refers to the total number of tweets the user posted), $log(NT/(ASR+1))$ (which is defined to represent the frequency of posting), $log(NFER+1)$ (where $NFER$ is defined as the number of the user's followers), $log(NFEE+1)$ (in which $NFEE$ denotes the number of the users' followees), $NT/(NFER+1)$, and $NT/(NFEE+1)$.
With respect to the privacy settings, corresponding binary features are compromised by whether a user allows comments from others, whether the user allows private messages sent from others and whether the user allows Weibo tracking its real-time locations. In addition, we consider the length of self-description as the feature.

\textbf{Interactive features} Interactive features are designed to reflect the sophisticated patterns of social interactions in Weibo at different time granularities of days or weeks. Here the social interaction includes posting, mentioning, and retweeting that have been verified to be key behaviors on the extraversion in the previous study. Specifically, for a certain time granularity, daily or weekly and a certain social interaction, a vector composed by averaged occurrences of the interaction (over the entire life of a user in our collection) at different hours or days of a week is first calculated and then from this vector, following features are extracted: (1) the average number of interactions, (2) the hour or day with the most interactions, (3) the maximum of occurrence of the hourly or daily interaction, (4) the hour or day with the least occurrence of the interaction, (5) the variance of the integration occurrence on different hours or days. Besides, the proportions of the tweets containing mentionings and retweetings are also considered as features to reflect the user's interactive intensity. 

\textbf{Linguistic features} Previous efforts on extraversion explorations have demonstrated that language styles in social media can be effectively indicators to infer personality traits. Because of this, we collect 261 terms that could describe the personality traits, including both Chinese and English, to linguistically model the tweets posted by users of different groups. After preprocessing the text, all tweets posted by a user is combined to form a document to represent the user's language style and all user's documents compose the corpus. Then the classic TF-IDF scores are employed to evaluate the 261 terms and the top 84 terms~\cite{Aizawa:2003in} are selected to extract linguistic features. Specifically, for any term within the 84 selected ones, if it occurs in a document (corresponding to a user) its feature value will be 1 otherwise 0. This method, called bag-of-word, is always utilized in natural language processing~\cite{Jiang:2007sys}. Meanwhile, we also consider the average length of tweets posted by the user. 

% normalizing
It is worth noting that in our dataset of online profiles, there are significant differences in the scale of the extracted features. In order to train an unbiased machine models, feature standardization is indeed a necessary requirement. We perform the standardization and transform each feature into the range between zero and one. The transformation is given by
\begin{equation}
 X_{i} = \frac{X_{i} - X_{min}}{X_{max} - X_{min}},
\end{equation}
where $X_{i}$ is the $i$-th item in the feature set $X$, $X_{max}$ is the maximal value of $X$, and $X_{min}$ is the minimal value of $X$.

In summary, we extract 130 features in total for each Weibo user, including 13 basic features, 32 interactive features and 85 linguistic features, which will be input of the machine learning models.

\subsection{Models and Accuracy}

Based on the training data and feature set obtained from the previous sections, three popular machine learning models, including Random Forest, Naive Bayes and Support Vector Machine (SVM) are employed to approach the 3-categories classification problem for extraversion. And regarding to the optimization algorithm of SVM, we choose C-SVM (multi-class classification) as the solution and RBF as the kernel function. We adapt 10-fold cross-validation to examine the average accuracy of different models. 

\begin{table}{}
\centering
\small
\caption{The average accuracy and F1-score of machine learning models.}
\begin{tabular}{llllll}
\hline
\multicolumn{1}{c|}{Model} & \multicolumn{1}{c}{Random Forest} & \multicolumn{1}{c}{Naive Bayes} & \multicolumn{1}{c}{\textbf{SVM}}     \\ \hline

\multicolumn{1}{c|}{Accuracy of inferring extroverts} & \multicolumn{1}{c}{36.98\%} & \multicolumn{1}{c}{47.11\%} & \multicolumn{1}{c}{\textbf{52.28\%}} \\ \hline

\multicolumn{1}{c|}{Accuracy of inferring introverts} & \multicolumn{1}{c}{37.75\%} & \multicolumn{1}{c}{46.44\%} & \multicolumn{1}{c}{\textbf{49.49\%}} \\ \hline

\multicolumn{1}{c|}{Accuracy} & \multicolumn{1}{c}{34.73\%} & \multicolumn{1}{c}{39.17\%} & \multicolumn{1}{c}{\textbf{42.31\%}} \\ \hline

\multicolumn{1}{c|}{F1-score} & \multicolumn{1}{c}{0.3933} & \multicolumn{1}{c}{0.4062} & \multicolumn{1}{c}{\textbf{0.4505}} \\ \hline

\label{accuracy}
\end{tabular}
\end{table}

The baseline of accuracy for 3-category classification is 33.33\%. As can be seen in Table~\ref{accuracy}, our 10-fold cross-validation results show that the Random Forest model cannot properly solve the classification of extraversion (with accuracy close to the baseline). The Naive Bayes and SVM models outperform the baseline solutions significantly, especially the SVM model, whose accuracy for both extroverts and introverts arrives around 50\%. In the meantime, we also measure the average F1-score by calculating the rate of precision and recall for each label $i$ and find that their unweighted mean that defined as
\begin{equation}
 F1\text{-}score = \frac{1}{3}\cdot\sum_{\substack{i=-1,0,1}}\frac{2 \cdot (precision_i \cdot recall_i)}{precision_i + recall_i}
\label{equ:F1}
\end{equation}
is 0.4505, indicating that not only on the rate of precision but also on recall the performance of SVM can be further justified to be convincing. Therefore, we train a SVM model to be the extraversion classifier, which can be employed later to identify extroverts and introverts in Weibo without the help of self-reports. Because of competent accuracy and $F1\text{-}score$, we argue that machine learning models like SVM can break the limitations of conventional approaches like self-reports and greatly extend the scope for personalty explorations and offer an opportunity to comprehensively picture the behavioral differences between extroverts and introverts in social media.

%---------------------------------------------------%
\section{Differences between extroverts and introverts}

Employing the obtained SVM classifier, we attempt to identify extroverts and introverts from a large population of Weibo users, whose online public available profiles were collected through Weibo's open APIs within the period between November 2014 and March 2016 and the ones with less than 100 tweets were omitted to avoid the sparsity. After converting each user into a representation of the feature set, our SVM classifier can automatically categorize it into an extrovert, neutral or introvert, respectively and from 16,856 users we totally get 4,920 extroverts and 2,329 introverts. The self-reports and online profiles of users mentioned in this study are publicly available to the research community after a careful anonymization, which can be downloaded freely through \url{https://doi.org/10.6084/m9.figshare.4765150.v1}. In order to establish a comprehensive spectrum of the behavioral discrepancy for extroverts and introverts in social media, patterns in perspectives of time, geography, online activities, emotion expressions and attitudes to virtual honor are systematically investigated according to our seven research questions.

\subsection{Temporal differences}
\label{sec:time}

\begin{figure}
\centering
\includegraphics[width=0.8\textwidth]{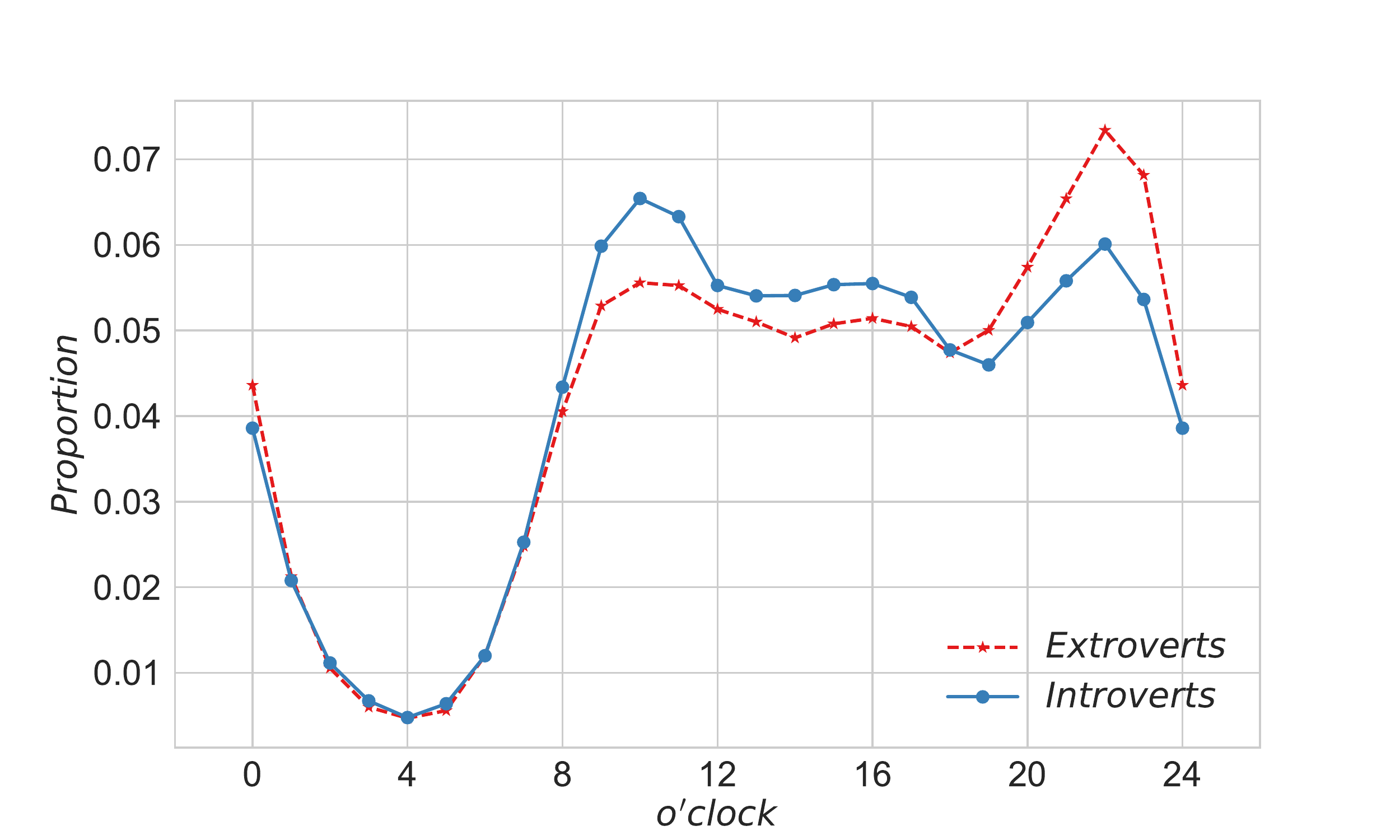}
\caption{Hourly pattern of the posting. The statistics of hourly proportions are obtained from all tweets posted by extroverts and introverts, respectively.}
\label{fig:time_series}
\end{figure}

Users of different personality traits might post tweets unevenly at different hours of a day, i.e., the hourly pattern of the posting and it can be reflected by the distribution of posted tweets at 24 hours of a day. As can be seen in Fig.~\ref{fig:time_series}, introverts prefer to post tweets from 8:00 to 18:00 while extroverts are active and excited from 19:00 to 1:00 of the next day, implying that extroverts are move vibrant than introverts at night. Further evidence can be found at the individual level in Table~\ref{tab:time_group} that the proportion of extroverts tweeting on daytime (from 8:00 to 19:00) is 0.557 and that of introverts is 0.608; the proportion of extroverts tweeting at night (from 19:00 to 1:00 of the next day) is 0.358 while that for introverts is 0.305. Active postings at night for extroverts suggests that their nightlife might be more diverse than that of introverts.

\begin{table}[]
\centering
\caption{Tweeting habits at different periods of a day at the individual level.}
\label{tab:time_group}
\begin{tabular}{@{}cccc@{}}
\toprule
Time       & 1:00 - 8:00   & 8:00 - 19:00  & 19:00 - 1:00 \\ \midrule
Extroverts & 0.085 & 0.557 & 0.358              \\
Introverts & 0.087 & 0.608 & 0.305              \\ \bottomrule
\end{tabular}
\end{table}

Posting intervals between two temporally consecutive tweets of an individual Weibo user can be an excellent indicator to reflect its degree of preference and dependency on the social media. We calculate the average interval (with unit of hour) for extroverts and introverts from timestamps of their tweets, respectively. As can be seen in Table~\ref{tab:time_interval}, introverts post their tweets more frequently than extroverts, implying heavier dependency on the social media. Specifically, the mean interval of introverts is 19.09 hours while that for extroverts is 28.10 hours. It is consistent and can be well explained by the previous finding that individuals who are in social isolation tend to depend on and indulge in the social media to relieve the loneliness due to lacking of interactions with others in the real life~\cite{DESARBO:1996ck}. Meanwhile, differences in standard deviation also reveals that introverts have more regularity in Weibo than extroverts from the perspective of posting frequency (respectively 74.36 and 62.25 hours). Moreover, if only considering the time interval within one day (namely ignoring the interval over 24 hours), the mean interval of extroverts shrinks to 6.41 hours and that of introverts is 5.61 hours. In this case, the standard deviation of time interval of extroverts is 6.76 hours and which is more than that of introverts. It further justifies the finding that introverts post more frequently than extroverts and demonstrates more significant preference on the social media usage.

\begin{table}[]
\centering
\caption{Average intervals and standard deviations of posting tweets for extroverts and introverts.}
\label{tab:time_interval}

\begin{tabular}{cccc}
\toprule
Time interval (hour) & Mean & Standard deviation \\ \midrule
Extroverts           & 28.10       & 74.36              \\
Introverts           & 19.09       & 62.25              \\ \bottomrule

Time interval (hour) within 24 hours   & Mean & Standard deviation  \\ \midrule
Extroverts           & 6.41        & 6.76               \\
Introverts           & 5.61        & 6.19               \\ \bottomrule
\end{tabular}
\end{table}

\subsection{Spatial differences}
\label{sec:geo}

An individual user can post geo-tagged tweets (or checkins) containing latitude and longitude of the location where the user is in Webo, which information indeed offers us a proxy to decently explore geographical differences between extroverts and introverts. To perform the geo-analysis, we extract 57,710 tweets with extract geographical locations, of which 38,729 tweets are posted by extroverts and 18,981 tweets by introverts.

\begin{figure}
\centering
\includegraphics[width=0.8\textwidth]{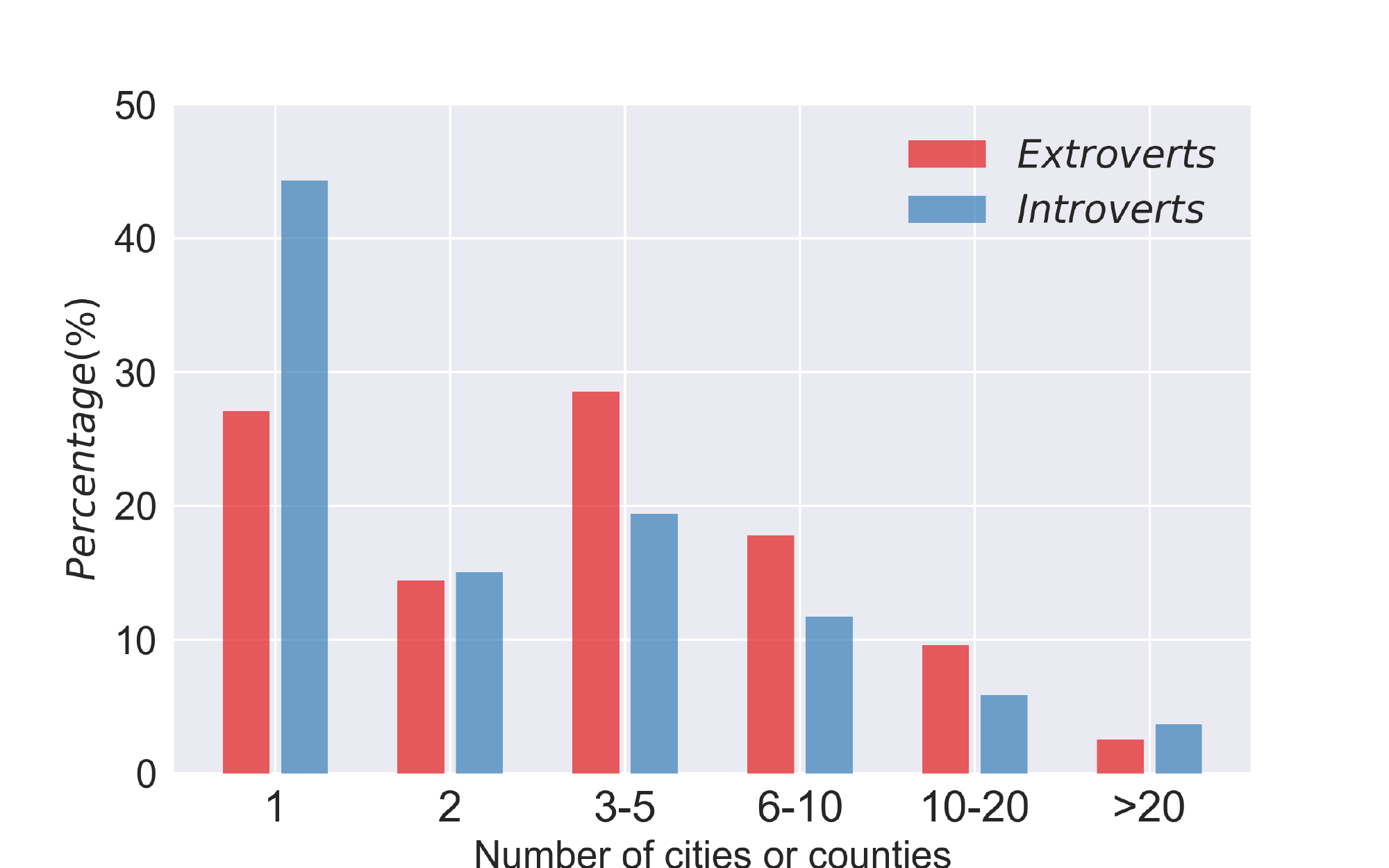}
\caption{Percentages of the number of posting tweets for extroverts and introverts, respectively.}
\label{fig:geo_city_bar}
\includegraphics[width=0.8\textwidth]{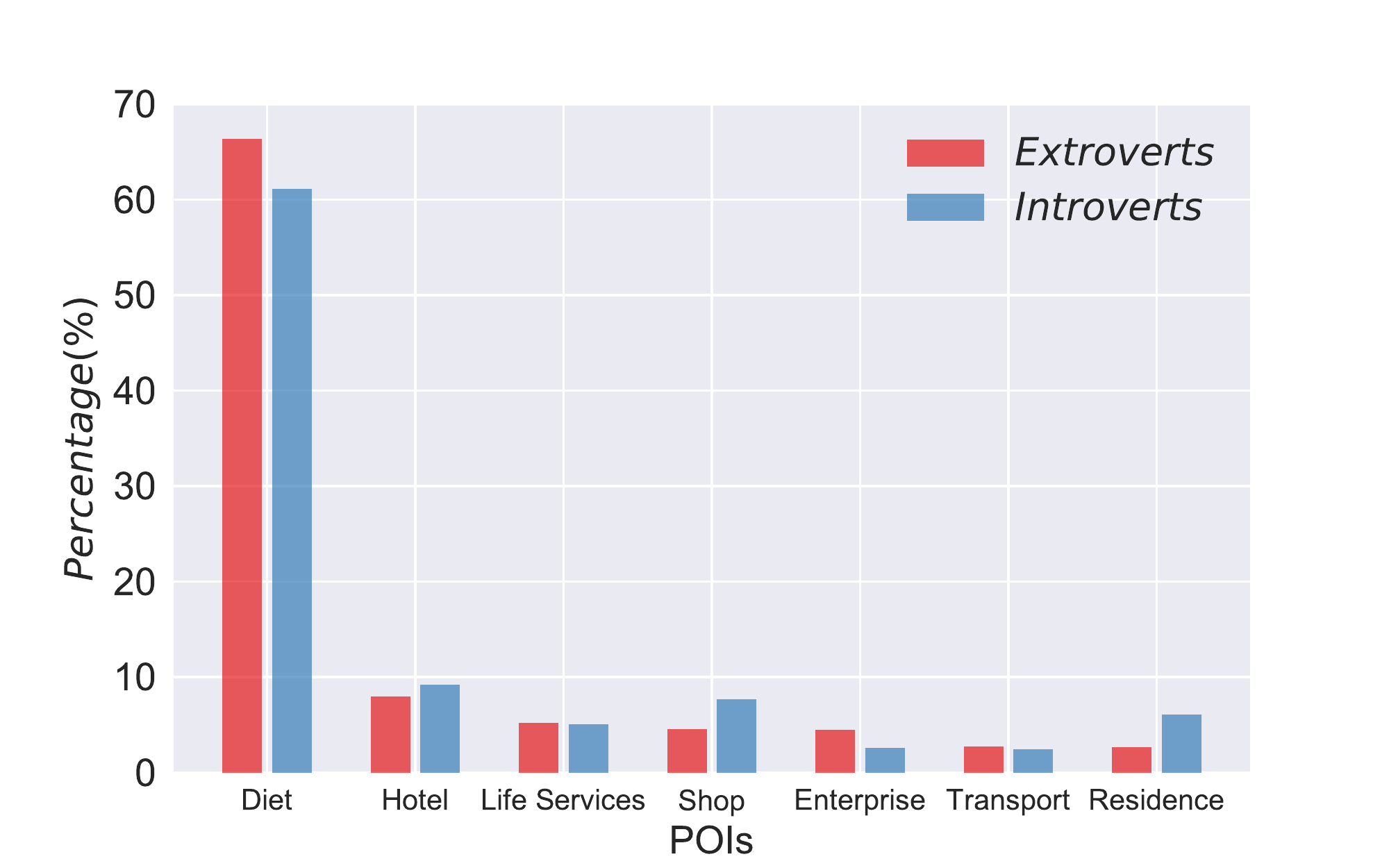}
\caption{Percentages of geo-tweets posted at different POIs by extroverts and introverts, respectively.}
\label{fig:POI_bar}
\end{figure}

For each geo-tagged tweet, we transform its longitude and latitude to the corresponding city (or county) through GeoPy project~\cite{Geopy:2013} and then for each user, we can accordingly obtain the list of cities or counties where it posted tweets. The results of the comparison between extroverts and introverts are shown in Fig.~\ref{fig:geo_city_bar} and which surprisingly demonstrates the significant differences in spatial life style of users with two kinds of personality traits. As can be seen, 44.32\% introverts post tweets only from one city or county, perhaps their residences, suggesting that nearly half of introverts prefer staying in just one familiar city or county. While as to extroverts, only 27.12\% of them are located in only one city or county, which is far less than that of introverts. To be more specific, 14\% of extroverts and 15\% of introverts post tweets in two cities (or counties), 29\% of extroverts and 19\% of introverts post tweets in 3-5 cities or counties and the trend of extroverts tweeting at more places holds persistently as the city (or county) number ranges from 6 to 20. It is noteworthy that the number of extroverts posting tweets in 3-5 cities is even more than the number located in one city, which implies that extroverts prefer going to or visiting more cites than introverts, in which posting at one city or county dominates. As the number of cities (or counties) is more than 20, it is unexpected that the ratio within introverts is significantly greater than that of extroverts, implying that a tiny fraction of introverts might attempt to camouflage their own loneliness to others by updating tweets with a large number of different places~\cite{DESARBO:1996ck}.

Beyond the city granularity, in fact we can also perform geographical comparison on better resolutions, like Point-of-Interest (POI), which is a detailed description of the featuring function for small regions or point locations of cities. Specifically, POIs exclude private facilities such as personal residences, but include many public facilities that seek to attract the general public such as retail businesses, amusement parks, industrial buildings and etc. Government buildings and significant natural features are POIs as well. They are also referred to hotels, restaurants, fuel stations or other categories in automotive navigation systems and recommendation systems~\cite{Ye:2011exploiting}. In this study, nine kinds of POIs referring to Restaurants, Hotels, Life services, Shops, Enterprises, Transportation, Entertainment, Neighborhoods and Education are considered and the percentages of the most six POIs visited by extroverts and introverts are respectively shown in Figure~\ref{fig:POI_bar}. It is found that most geo-tagged tweets are posted from restaurants, occupying 66.38\% for extroverts and 61.10\% for introverts within the nine kinds of POIs we select. There are significant differences between extroverts and introverts on visiting Shops and Enterprises. The percentage of tweets located in shops is 4.58\% for extroverts while 7.68\% for introverts, implying that introverts prefer to checkin or post tweets while shopping as compared to extroverts. Even more interesting, the percentage of tweets located in companies and enterprises is 4.46\% for extroverts and 2.59\% for introverts. Since companies and enterprises are always the working places of individuals, it is suggested that extroverts tend to inform others as they are working.

\subsection{Online activities}
\label{sec:behavior}

Diverse online activities, like sharing, interacting and buying in Weibo can be exactly sensed only through the tweets posted by users. Hence by ming texts of tweets posted by extroverts and introverts, we try to offer a behavioral difference landscape of online activities. 

\begin{figure}
\centering
\includegraphics[width=0.8\textwidth]{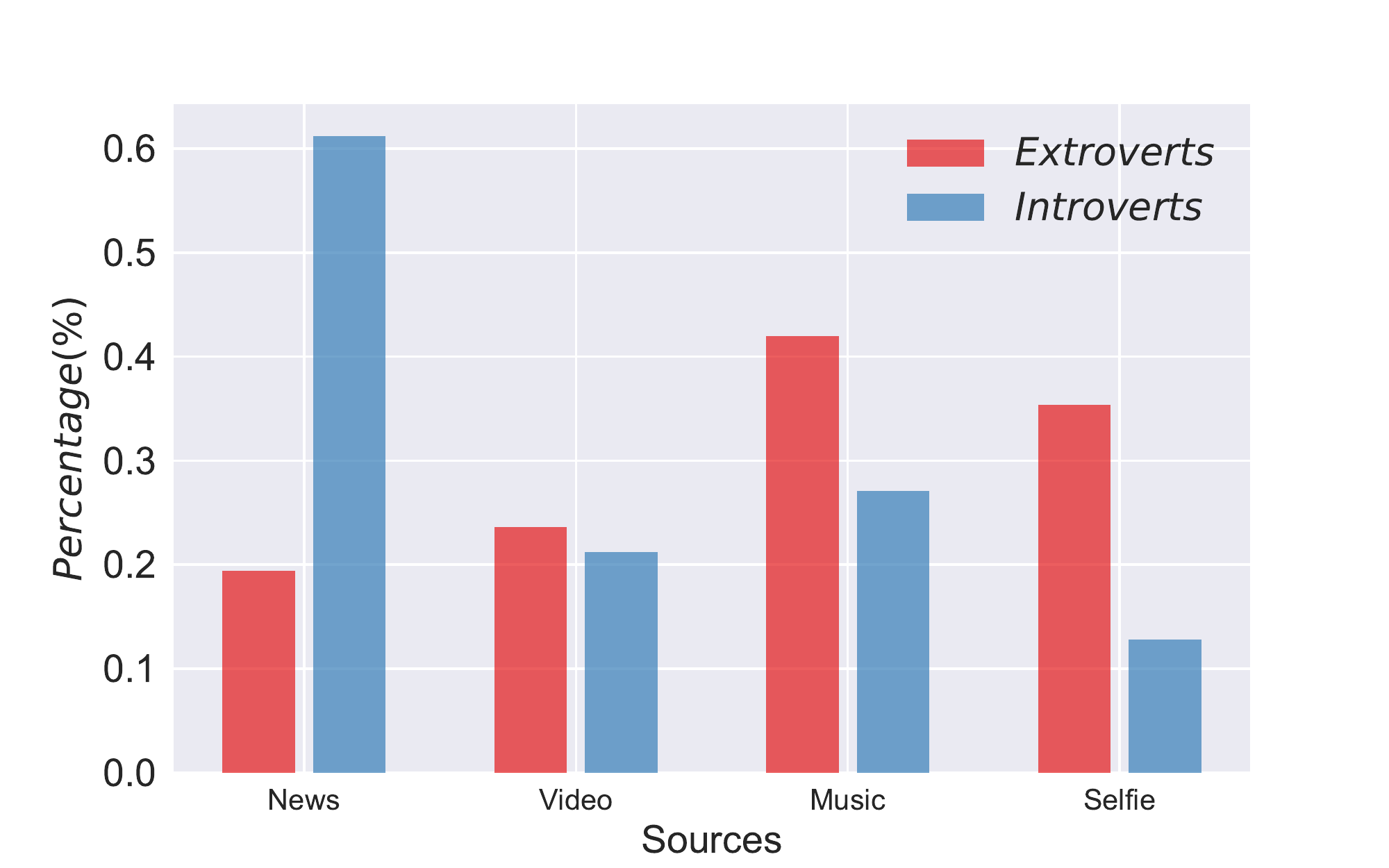}
\caption{Percentages of four kinds of sharing by extroverts and introverts in Weibo.}
\label{fig:source_bar}
\end{figure}

\textbf{Sharing} Each tweet in Weibo is labeled by a tag to manifest its posting source. For example, if a user logs in Weibo and posts one tweet, the source could be mobile devices (e.g. iPhone) or web browsers (e.g. Chrome). In the meantime, Weibo users always share news, videos, music to their friends or the public in social media and diverse sources of these shared information are also kept in tweets posted in terms of news websites, mobile applications or other social platforms which offer the sharing interface to Weibo. Besides, tweets shared from the selfie mobile softwares will also be tagged as selfies, which always contain the self-portrait photograph typically taken by the camera phone. Because of these features, in this study we utilize the source label of each tweet to analyze the sharing behavior of extroverts and introverts. The occupations of the above four sharing in all tweets are demonstrated in Figure~\ref{fig:source_bar}. As can be seen, the fraction of news sharing of introverts (0.6122\%) is three more times than that of extroverts (0.1939\%), contrarily, extroverts enjoy sharing more videos, music and selfies in social media than introverts, especially the selfies, e.g., the fraction of selfie tweets for extroverts is 0.3539\% and is much higher introverts' 0.1276\%. It is widely believed that selfie is connected with individual narcissism~\cite{sorokowski2015selfie,weiser2015me,wang2017study} and our findings further suggest that extraversion is positively coupled with selfie in social media. 

\textbf{Interacting} Interacting patterns, especially the mentioning and retweeting are actually considered comprehensively in the feature set we extract in section ~\ref{sec:clas} and which are then used be the input of the extraversion classifier. Intuitively, performing the analysis of behavioral difference of interacting patterns on extroverts and introverts identified by the classifier would be meaningless, because differences have already been latently considered in the classifier. In order avoid the biased comparison and provide solid evidence, here we perform the difference analysis directly on the training set, i.e., users filled the self-reports. In terms of Pearson correlation, we measure the linear dependence between the interaction features and the extraversion scores of participants. As can be seen in Table~\ref{tab:interactions}, features with the relatively high Pearson correlations (coef.$>0.13$) with extraversion scores are listed in details. It's interesting to find that the features related to the mentioning behavior (@ behavior) are positively correlated with extraversion scores. The mentioning behavior is regarded as one of the most important form of online interactions. Specifically, both rate of @ in all tweets and the average number on posting most tweets with @ within one hour in one day can evidently reflect the frequency of interaction with other users of the individual. Meanwhile, the variance of posting tweets with @ on hours of a day or on days in one week suggest the irregularity and randomness of the mentioning behavior of users. Therefore from Table~\ref{tab:interactions} we conclude that the extroverts are more socially active and interacting than the introvert in social media, however, their interactions are more casual and temporally less regular than that of introverts.

\begin{table}[]
\centering
\caption{Interacting features with relatively high Pearson correlations with extraversion scores.}
\begin{tabular}{@{}ll@{}}
\toprule
Feature                                                   & coef.                \\ \midrule
Rate of @ in tweets                                       & 0.172                \\
Average number of posting most tweets with @ within one hour & 0.150                \\
Variance of posting tweets with @ on hours in one day    & 0.135                \\
Variance of posting tweets with @ on days in one week    & 0.131                \\ 
\bottomrule
\label{tab:interactions} 
\end{tabular}
\end{table}

\begin{figure}
\centering
\includegraphics[width=0.6\textwidth]{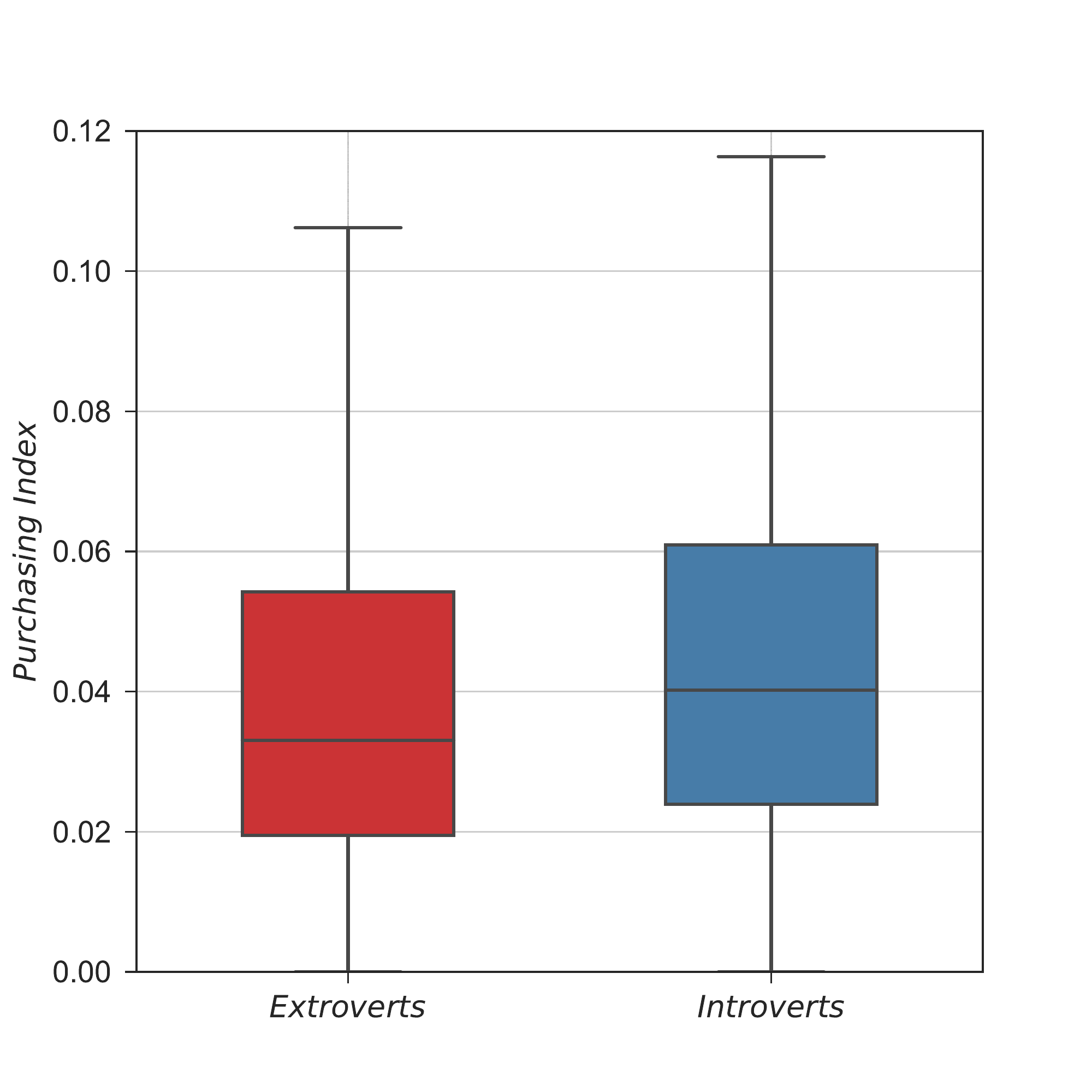}
\caption{Box-plot of the Purchasing Index of extroverts and introverts. The bottom line of the box represents the 25th percentile, the line inside the box represents the median, the uppermost line of the box represents the 75th percentile, and the topmost vertical line represents the maximum of the Purchasing Index.}
\label{fig:purchase_box}
\end{figure}

\begin{figure}
\centering
\includegraphics[width=0.6\textwidth]{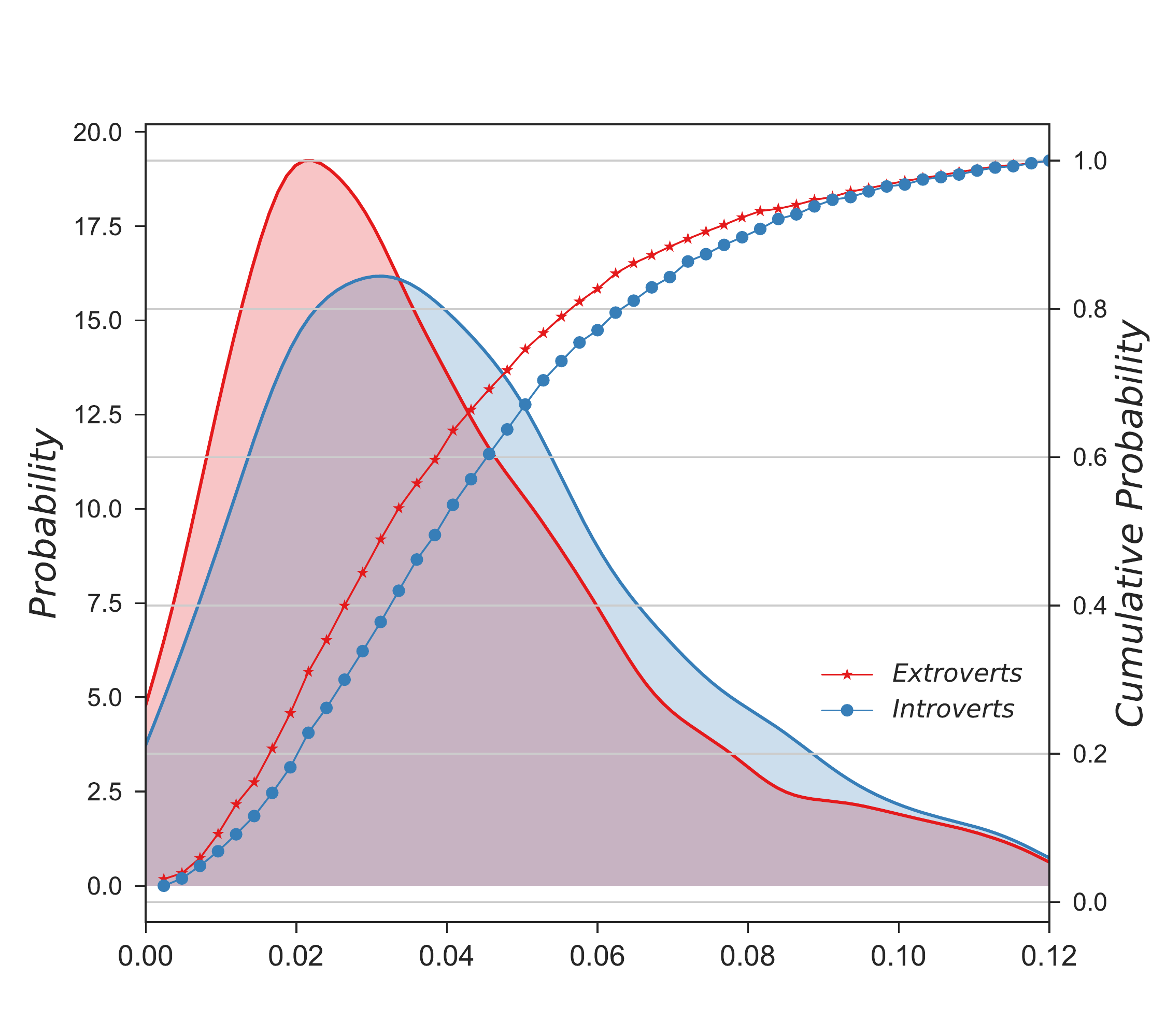}
\caption{The probability distribution of the Purchasing Index of extroverts and introverts.}
\label{fig:purchase_cdf}
\end{figure}

\textbf{Buying} The most intrinsic nature of social media is updating users' every status to their friends and thus experience like buying or shopping can be accordingly sensed through counting related keywords, namely the word-count method and which has been employed extensively in the field of psychology~\cite{kramer2010unobtrusive} in recent decades. In this study, 14 buying keywords are selected to sense the buying behavior (like BUY and SHOPPING), response in sales promotion (like DISCOUNT and 11.11, a famous day for promotion sales in China advocated by Taobao inc.) and mentioning or sharing of online shopping malls (like AMAZON and TAOBAO). For tweets posted by extroverts or introverts, the ones containing one or more selected keywords will be labeled as buying related and the fraction of buying tweets of each user is defined as Purchasing Index, reflecting its intensity of the buying behavior or purchasing intention. By calculating the Purchasing Index of each user, the comparison between extroverts and introverts on buying behavior is depicted in Fig.~\ref{fig:purchase_box}. The mean of Purchasing Index of extroverts is 0.0440 and that of introverts is 0.0484, which is 10\% larger than that of extroverts. The 25th percentile, the median, the 75 percentile and maximum of Purchasing Index of extroverts are respectively 0.0199, 0.0331, 0.0543 and 0.7736, and those of introverts are 0.0239, 0.0402, 0.0609 and 0.8480. Fig.~\ref{fig:purchase_cdf} further shows the commutative distribution function (CDF) and probability distribution function (PDF) of the Purchasing Index of extroverts and introverts. As can be seen, the Purchasing Indexes of more than 95\% of users are smaller than 0.1 and the significant difference between extroverts and introverts also mainly locates in this region. Specifically, at the same Purchasing Index level (like $>0,>0.04,>0.06,>0.08$), the probability of introverts is always greater than extroverts, surprisingly suggesting that introverts prefer to publish tweets referring to purchasing than extroverts. This conclusion could apply to the advertising and sales of commodity and other realistic scenarios, i.e., introverts might be ideal marketing targets in online promotions. In addition, we perform the analysis of variance (ANOVA) to further testify the results, which method is used to analyze the differences among group means and their associated procedures. We adapt it to investigate whether there are significant differences of Purchasing Index between the two groups we discussed. As can be seen Table.~\ref{anova_purchase_index}, the $p$-value is less than 0.001 and the differences of two groups are statistically significant.

\begin{table}[]
\small
\centering
\caption{ANOVA on Purchasing Indexes.}
\label{anova_purchase_index}
\begin{tabular}{@{}cccccc@{}}
\toprule
               & Sum of Squares & Df   & Mean Square & F      & $p$-value. \\ \midrule
Between Groups & .029           & 1    & .029        &        &             \\
Within Groups  & 14.563         & 7247 & .002        & 14.497 &  $1.4*10^{-4}$    \\
Total          & 14.593         & 7248 &             &        &             \\ \bottomrule
\end{tabular}
\end{table}

\subsection{Online emotion expressions}
\label{sec:emotion_honor}

Tweets in social media not only deliver the factual information but also feelings of users and these feelings can be automatically identified into different emotions by mining only texts of tweets~\cite{Zhao:2012jc}. Because it is widely believed that extraversion is associated with higher positive affect, namely, extroverts may experience more positive emotions~\cite{mccrae2003personality,Smillie:2015fl}, thus in this study we try to investigate the differences between extroverts and introverts from the perspective of emotion expressions. By employing a previously built system named MoodLens~\cite{Zhao:2012jc}, we can categorize each tweet into one of five emotions, including anger, disgust, happiness, sadness and fear. Note that the tweets without significant emotional propensity will be ignored. Then for each individual, either extrovert or introvert, we calculate its emotion index for all five sentiments, which is defined as the fraction the corresponding emotional tweets in its tweeting history and quantificationally represents its emotional disposition in social media. 

\begin{figure}[]
\centering
\subfloat[$Anger\ Index$]{\includegraphics[width=0.45\textwidth]{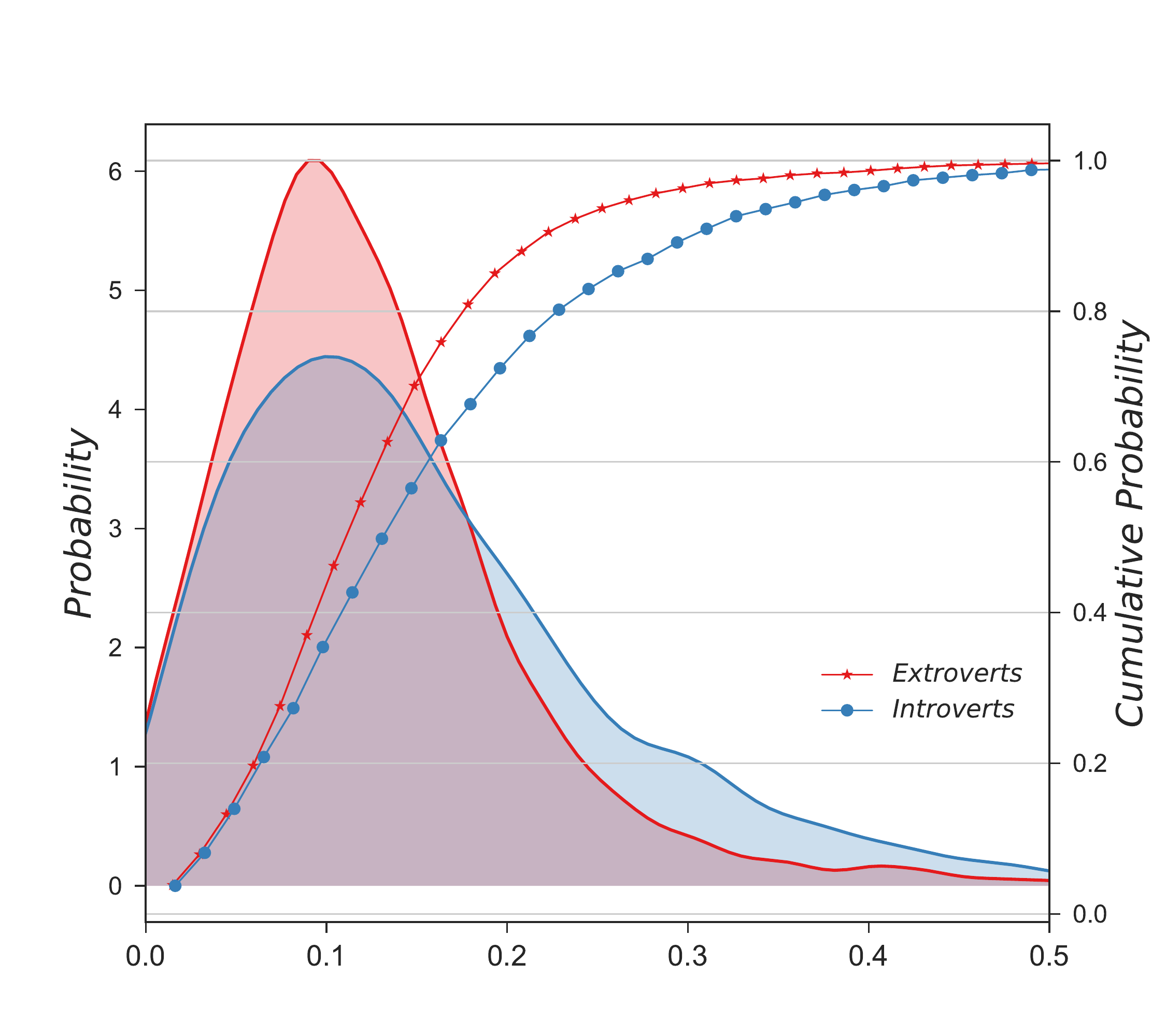}
\label{fig:anger}}
\hfil
\subfloat[$Fear\ Index$]{\includegraphics[width=0.45\textwidth]{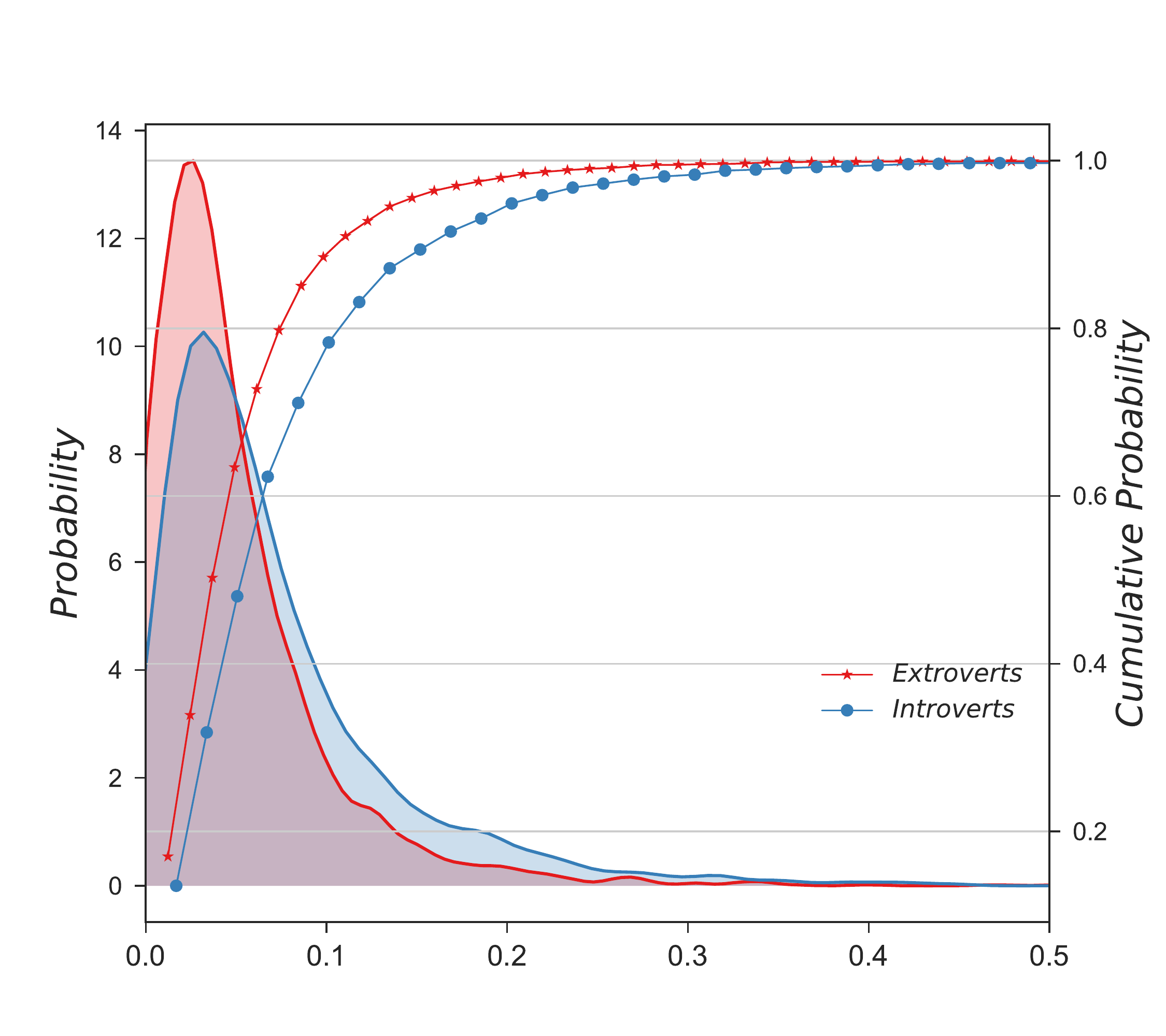}
\label{fig:fear}}
\hfil
\subfloat[$Sadness\ Index$]{\includegraphics[width=0.45\textwidth]{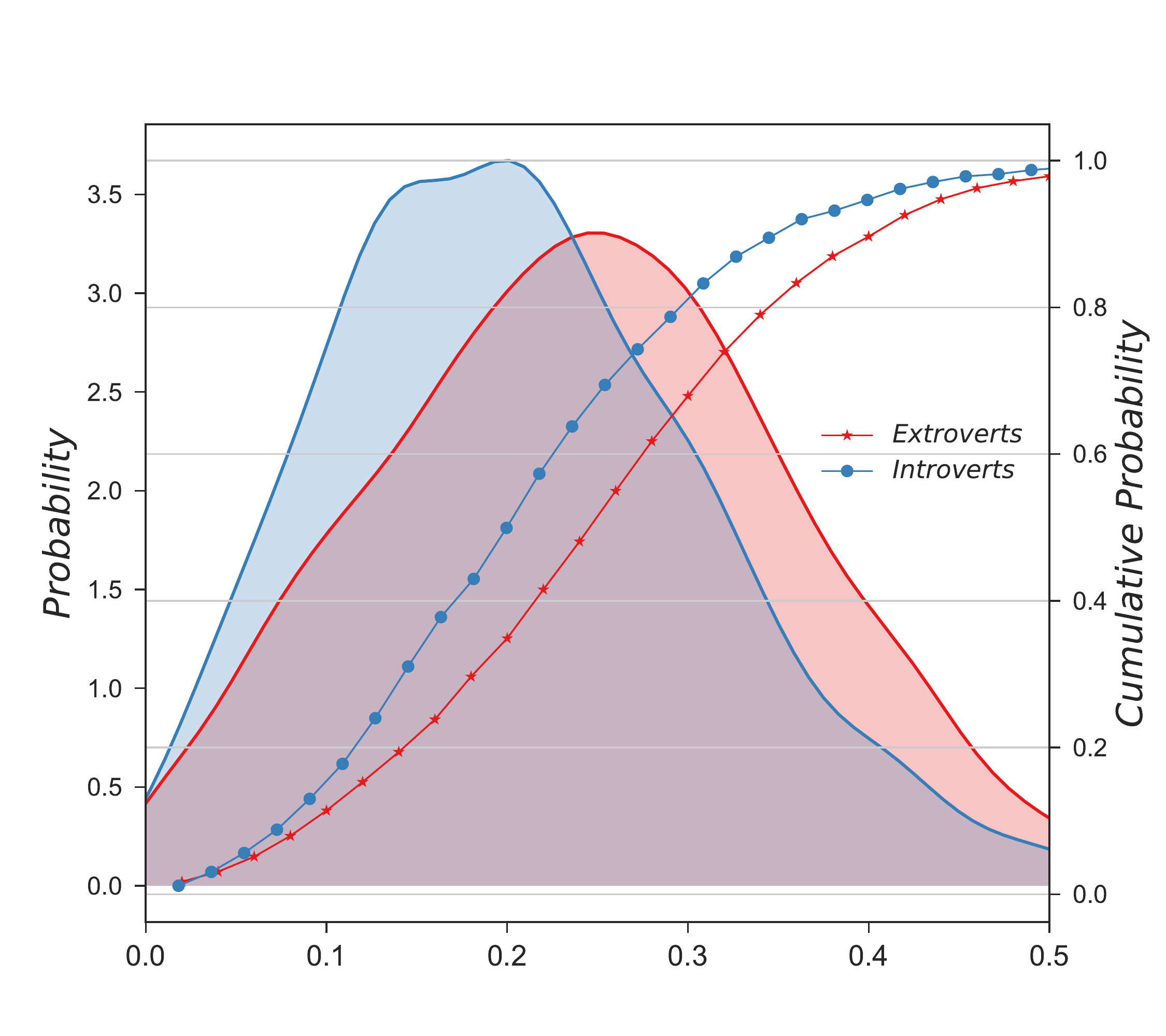}
\label{fig:happiness}}
\hfil
\subfloat[$Happiness\ Index$]{\includegraphics[width=0.45\textwidth]{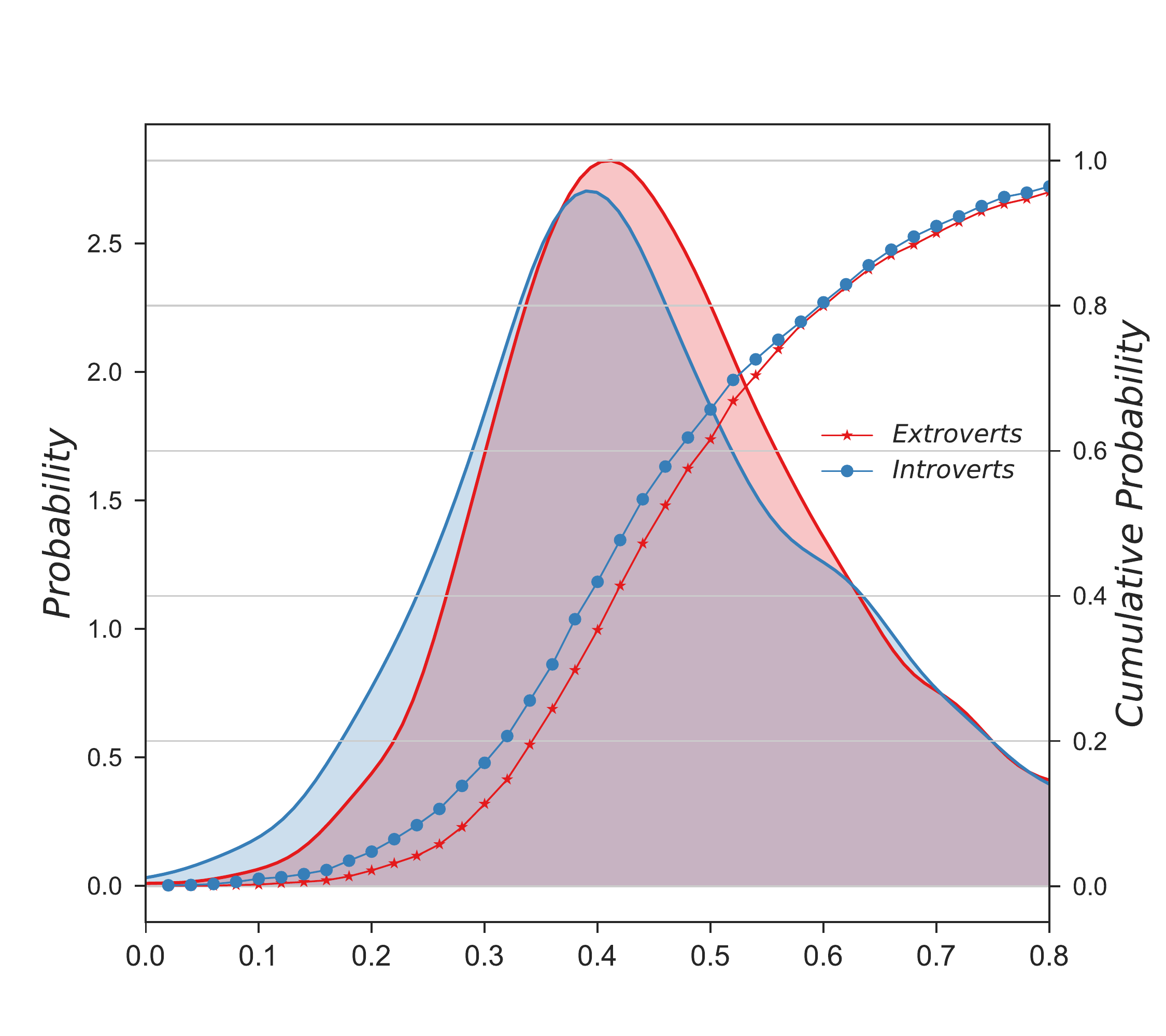}
\label{fig:sadness}}
\hfil
\subfloat[$Disgust\ Index$]{\includegraphics[width=0.45\textwidth]{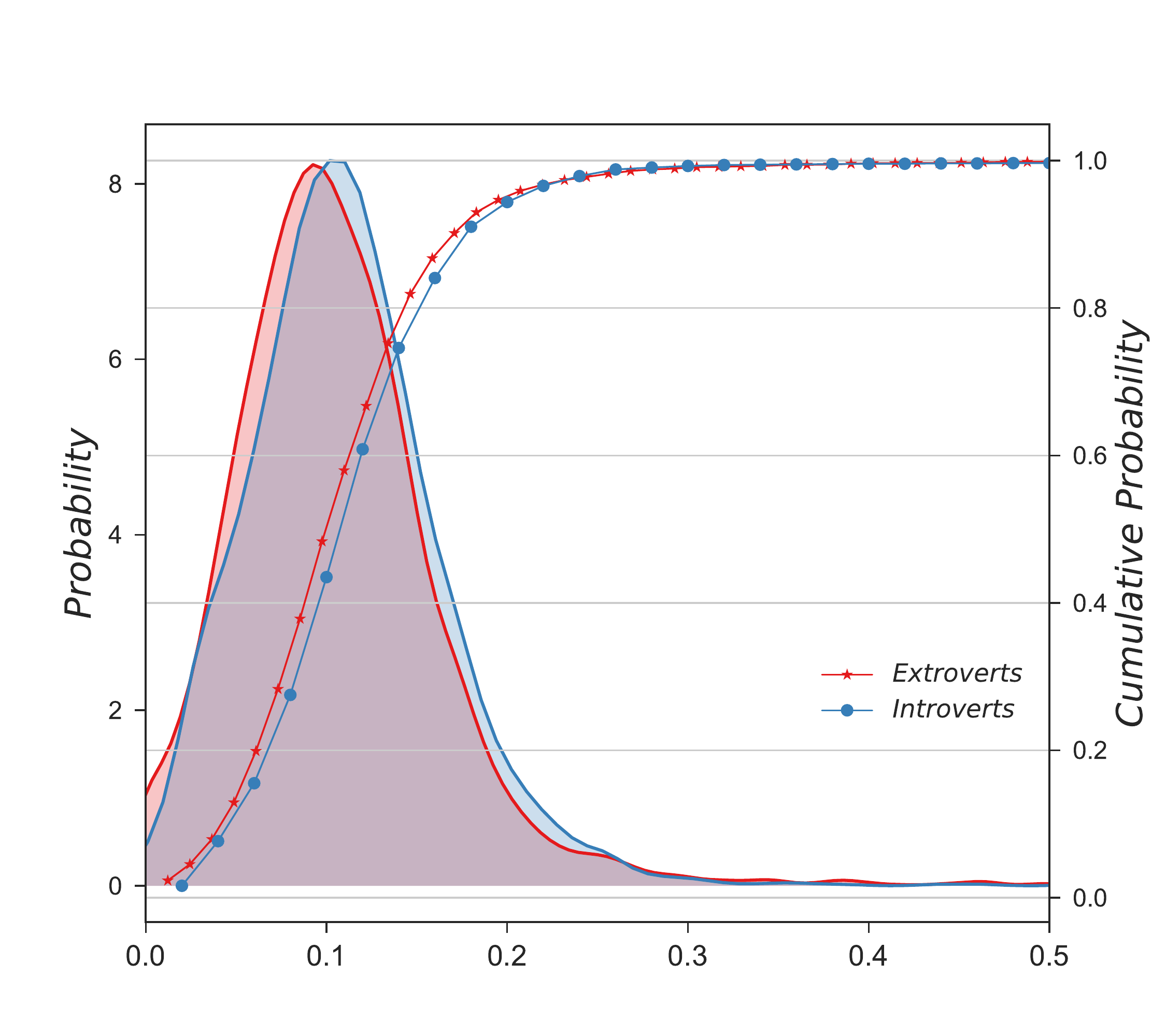}
\label{fig:disgust}}
\caption{CDFs and PDFs of five emotion indexes for extroverts and introverts, respectively. The mean values for Anger Index are 0.1237 and 0.1545, for Fear Index are 0.0496 and 0.0726, for Sadness Index are 0.2474 and 0.2084, for Happiness Index are 0.4737 and 0.4515 and for Disgust Index are 0.1056 and 0.1130.}
\label{fig:emotion}
\end{figure}

Fig.~\ref{fig:emotion} demonstrates CDFs and PDFs of five emotion indexes of extroverts and introverts, respectively. As can be seen, at the same Anger Index level ($(0.1, 0.4)$), Fear Index level ($(0, 0.25)$) and Disgust Index level ($(0.05, 0.15)$), the probabilities of introverts are always larger than that of extroverts, implying that introverts post more tweets of negative feelings than extroverts. However, for Sadness Index and Happiness Index, the probabilities of introverts are always less than extroverts, suggesting that extroverts tweet joy or sadness with more likelihood. Note that as can be seen in Figs.~\ref{fig:happiness} and \ref{fig:disgust}, the difference on Happiness Index and Disgust Index might be subtle but the Welch's test testifies the significance with $p\text{-value}<0.001$. Our finding is consistent with the previous statement that extraversion is associated with higher positive affect, however, in the meantime we also offer evidence that introversion is associated with the high arousal and negative affections like anger, fear and disgust and extraversion is positively correlated with sadness. Indeed with the help of data-driven approaches on large samples, our study can testify the existing conclusion and gain new insights at the same time.

\subsection{Attitudes to virtual honor}
\label{sec:honor}

\begin{figure}
\centering
\includegraphics[width=0.5\textwidth]{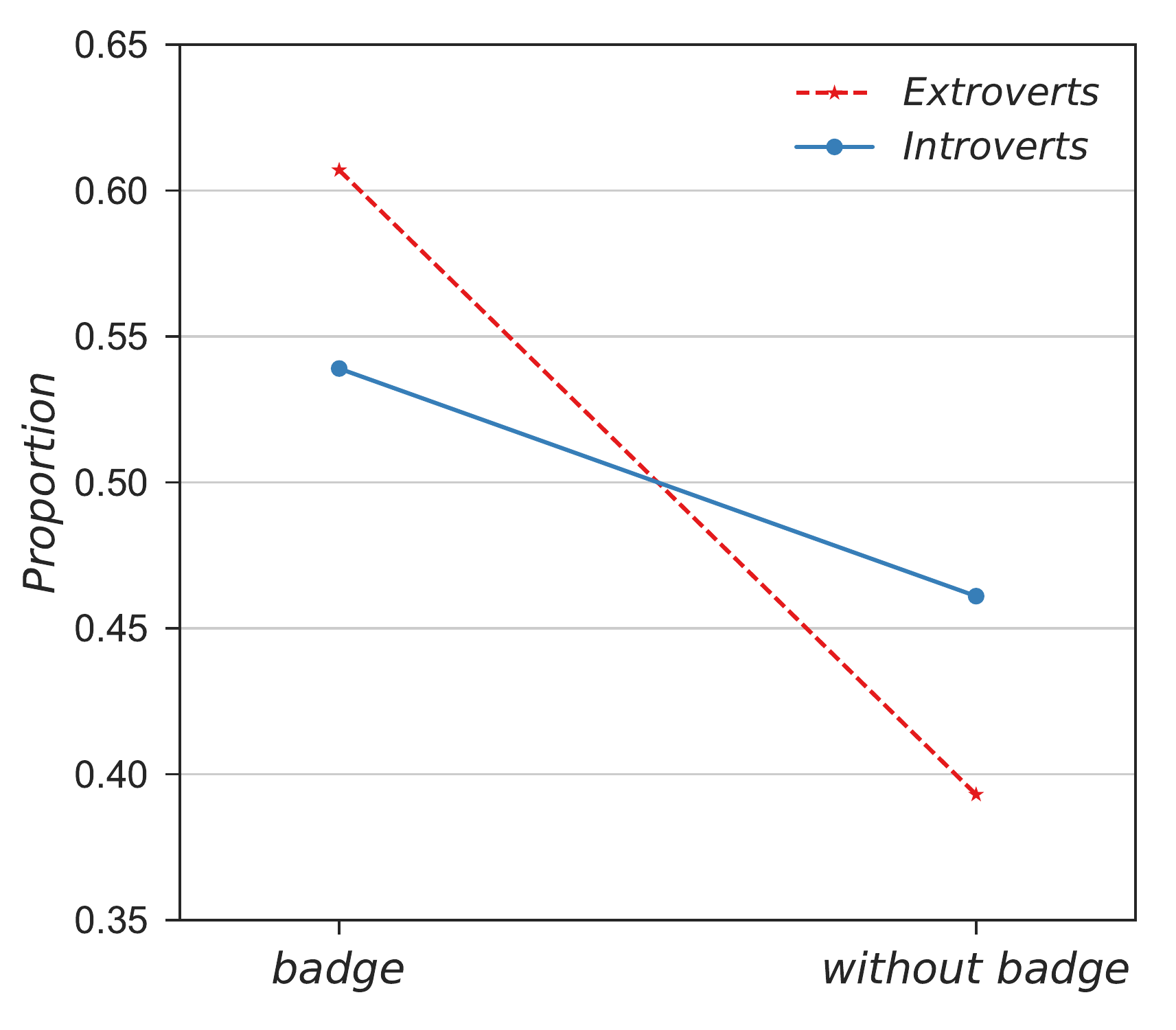}
\caption{The proportions of with and without the badge of Binding-Taobao for extroverts and introverts. }
\label{fig:badge}
\end{figure}

Weibo grants many optional badges to users, which they have to ``lighten'' through finishing necessary operations following demand of the social media. For instance, the users should connect the Weibo account and the Taobao account if they would like to get the ``Binding-Taobao'' badge. This behavior, exposing the Taobao account to the social media, is a risk of property security and privacy. However, the badges that users obtained are displayed publicly to the others and treated as honor in the virtual world. Because of this, a user's response to badges can be an indicator of its attitude to the virtual honor in social media. Then we investigate the difference of attitudes to virtual honor of extroverts and introverts, respectively. The ``Binding-Taobao'' badge is regarded as one relevant badge to perform the difference analysis and the distribution between extroverts and introverts of ``Binding-Taobao'' badges is shown in Fig.~\ref{fig:badge}. The percentage of extroverts with the ``Binding-Taobao'' badge is 60.7\% and that without the badge is 39.3\%. The percentage of introverts with the ``Binding-Taobao'' badge is 53.9\% and that without the badge is 46.1\%. It's obvious that the proportion of extroverts who obtain ``Binding-Taobao'' badges is larger than that of introverts. Besides, we also examine other various badges in Weibo, including ``Red envelope 2015'', ``Public welfare'', ``Travel 2013'', ``Red envelope 2014'' and etc. All the statistics of the badge indicate that extroverts tend to prefer badges than introverts do, in other words, extroverts attach more importance to the online virtual honor.

To sum up, from perspectives of tempo-spatial patterns, online activities, emotion expressions and attitudes to virtual honor, we establish a comprehensive picture of how extroverts tweet differently from introverts in social media.

\section{Conclusion}

Personality traits, like extraversion, are believed to play fundamental roles in driving human actions, however, a detailed and comprehensive understanding of how people with different personality traits behave is still missing, especially in the circumstance of the social media, which has been becoming an indispensable part of our daily life to date. Meanwhile, the lacking of large samples and the unavoidable subjectivity mean the conventional manners like self-reports may bring about bias on this issue. Hence in this study, we argue that starting from a small-scale but refined voluntary samples, establishing a map between self-reports and online profiles can help train a machine learning model to automatically infer massive individuals' personalities objectively without the costly expense on survey questionnaires. Indeed, the SVM classifier help us filter out over 7,000 extroverts and introverts from Weibo and to our best knowledge, build the first complete picture of how extroverts and introverts tweet differently in social media from perspectives like tempo-spatial patterns, online activities, emotion expressions and attitudes to virtual honor. Not only obtaining consistent conclusions with existing statements from traditional ways, new and insightful conclusions are also systematically revealed. Our findings offer solid evidence to the feasibility of machine learning based approach in personality research and will shed lights on realistic applications like online marketing and behavior understandings.

This study has inevitable limitations. For example, according to the Big-Five model, the individual personality also possesses other traits like openness, conscientiousness and so on, which will be promising directions in our future work.

\section*{Acknowledgments}
This work was supported by NSFC (Grant No. 71501005) and the fund of the State Key Lab of Software Development Environment (Grant Nos. SKLSDE-2015ZX-05 and SKLSDE-2015ZX-28).

\section*{References}
% \bibliography{mybib}

\end{document}